\documentclass[10pt,journal,compsoc]{IEEEtran}
\usepackage{graphicx}
\usepackage{url}
\usepackage{enumitem}
\usepackage{xcolor}
\usepackage{caption,subcaption}

\begin{document}
\title{The Security of IP-based Video Surveillance Systems}
%
%


\author{Naor~Kalbo\IEEEauthorrefmark{2},
	Yisroel~Mirsky\IEEEauthorrefmark{2},
	Asaf~Shabtai,
	and~Yuval~Elovici
	\IEEEcompsocitemizethanks{\IEEEcompsocthanksitem The authors are with the Department
		of Software and Information Systems Engineering, Ben-Gurion University, Be'er Sheva. Yisroel Mirsky is also with the Georgia Institute of Technology, College of Computing, USA.\protect\\
		E-mail: \{kalbo, yisroel\}@post.bgu.ac.il and \{shabtaia, elovici\}@bgu.ac.il\protect\\
\IEEEauthorrefmark{2} These authors contributed equally}
	\thanks{Manuscript received * **, ****.}}



%
\maketitle              

\begin{abstract}
IP-based Surveillance systems protect industrial facilities, railways, gas stations, and even one's own home. Therefore, unauthorized access to these systems has serious security implications. In this survey, we analyze the system’s (1) threat agents, (2) attack goals, (3) practical attacks, (4) possible attack outcomes, and (5) provide example attack vectors.



\end{abstract}

\section{Introduction}\label{sec:Intro}
\IEEEPARstart{T}{hese} days, video surveillance systems can be found everywhere. They are in the streets, at train stations, workplaces, factories, and even at home. Intelligent applications have made large surveillance networks practical to manage an utilize. For example, technology for facial recognition, identifying threats, event-detection, tracking objects, and rapidly investigating incidents, can be scaled to thousands of cameras over large geographical areas.

Over the last few decades, surveillance technologies have evolved from analog systems to packet switched systems (over IPv4 \& IPv6 networks). Moreover, video surveillance systems have become affordable due to the popular and pervasive Internet of Things (IoT). As a result, the market for security devices in connected homes has grown by a factor of 17 over the last few years \cite{statistacams}. Due to their convince, practicality, and affordability, video surveillance systems have become ubiquitous in our daily lives.

Unfortunately, in recent years, these systems and their components have been the target of cyber attacks. For example, they have been the target of distributed denial of service (DDoS) attacks,  exploited to invade the users' privacy, and even to mine cyrpto-currency. These systems have also been recruited into botnets to perform nefarious tasks. For example, in 2014, the infamous Mira botnet targeted surveillance systems, and infected over 600,000 devices worldwide \cite{antonakakis2017understanding}.

Through \url{Shodan.io} and {Censys.io} queries of well known manufacturers, we found over 1 million surveillance cameras and over 125 thousand surveillance servers exposed to the Internet. Of these devices, 90\% do not have secure login portals (use HTTP and not HTTPS). Moreover, approximately 8\% have open SSH and Telnet ports, 3\% have exposed MySQL databases, and at least 1.7\% of these devices are still vulnerable to the HeartBleed SSL vulnerability discovered in 2012. Even large video surveillance manufacturers have exposed products. For example, Samsung's CCTV Server has at least 83,035 exposed devices, where 86\% of them use HTTP login portals, and 1,604 have ssh ports open. Moreover, HikVision, the surveillance manufacture with the largest market share of 24.7\% has at least 260,415 exposed devices where only 53 of them had HTTPS enabled, but with self-signed certificates. These statistics emphasize the poor state of security of IP-based surveillance systems. 

In this article, we will review the cyber security of modern surveillance systems. We will start by detailing the composition and topology of modern video surveillance systems. Next will understand the goals of an attacker by discussing them in terms of their affect on the confidentiality, integrity, and availability of the system. Afterwards, we will explore how an attacker can realize his/her goal through multiple attack steps involving different threat agents and malicious actions. We will exemplify these attacks will current events and published common vulnerability exposures (CVE). Finally, we will review best practices and known security solutions which can be used to help mitigate these cyber threats.

We note that many other works have performed comprehensive reviews of IoT security \cite{rose2015internet, ling2017end, khan2018iot}. Although, IP-based surveillance systems use similar technologies, they are different with regards to cyber security. This is because surveillance systems are cyber-physical systems: they virtually support and enforce our physical security. When compromised, there is a threat to our physical safety, at home, at work, or in a national sense. As a result, the technology, threat actors, and attack motivations differ from other IoTs:
\begin{description}
    \item [Threat Actors (who)] There are attackers which want to exploit the functionality of these systems specifically. For example, state-actors or thieves performing reconnaissance over a geographic area and criminals planning to blackmail a victim with video footage. 
    \item [Assets (what)] If compromised these systems can provide an attacker with private imagery resulting in a direct explicit violation of privacy. These systems are also lucrative assets to botnet owners since they typically have high bandwidth (for DDoS attacks) and decent compute capabilities (for cryptomining).
    \item [Topology (where)] Unlike other IoTs, surveillance systems are often centralized systems connected to a single server. They are also commonly connected to both the Internet and an internal private networks --thus exposing a potential infiltration vector. 
    \item [Motivation (why)] Aside from being a stepping stone into another network, surveillance systems elicit monetary motivations such as blackmail, cryptomining, and spying for military or political reasons. Moreover, an attacker can have a physical advantage if the system is targeted in a DoS attack. For example, stopping video-feeds in certain geographic areas prior to an attack/theft, or as an act of cyber terrorism.
    \item [Attack Vectors (how)] These systems have often unique security flaws due to their functionality. For example, surveillance video servers often try to be open-platform and compatible with many different camera models. As a result, these servers often use obsolete encryption suites, and accept self-signed certificates which can lead to man-in-the-middle attacks. Moreover, modern systems rely on machine learning algorithms to identify and track objects and people. Unlike AI on other IoTs, these AI models can easily evaded/exploited due to their accessibly and flaws~\cite{akhtar2018threat}.
\end{description}

We also note that here is a good security review of CCTV and Video Surveillance Systems in \cite{costin2016security}. However, in \cite{costin2016security} the authors focus on visual attacks such as data exfiltration, covert channels, and steganography. In this survey, we provide the reader with a systematic review of the security of modern surveillance systems. We provide a comprehensive enumeration and description of vulnerabilities and attacks which pertain to surveillance systems. We hope that this systematic survey will provide the reader with a good understanding of how these systems are and can be exploited. We also discuss emerging threats and future research direction. 
\color{black}

\section{System Overview}\label{sec:IPsurvSys}
Before we can discuss the security aspects, we must describe what a surveillance system is. In this section, we first present a general overview of IP-based video surveillance systems. Afterwards, we list some of the system's critical assets and common deployment schemes. 

\subsection{Overview}\label{sucsec:taxonomy}
\begin{figure*}[!ht]
	\centering
	\includegraphics[width=.9\textwidth]{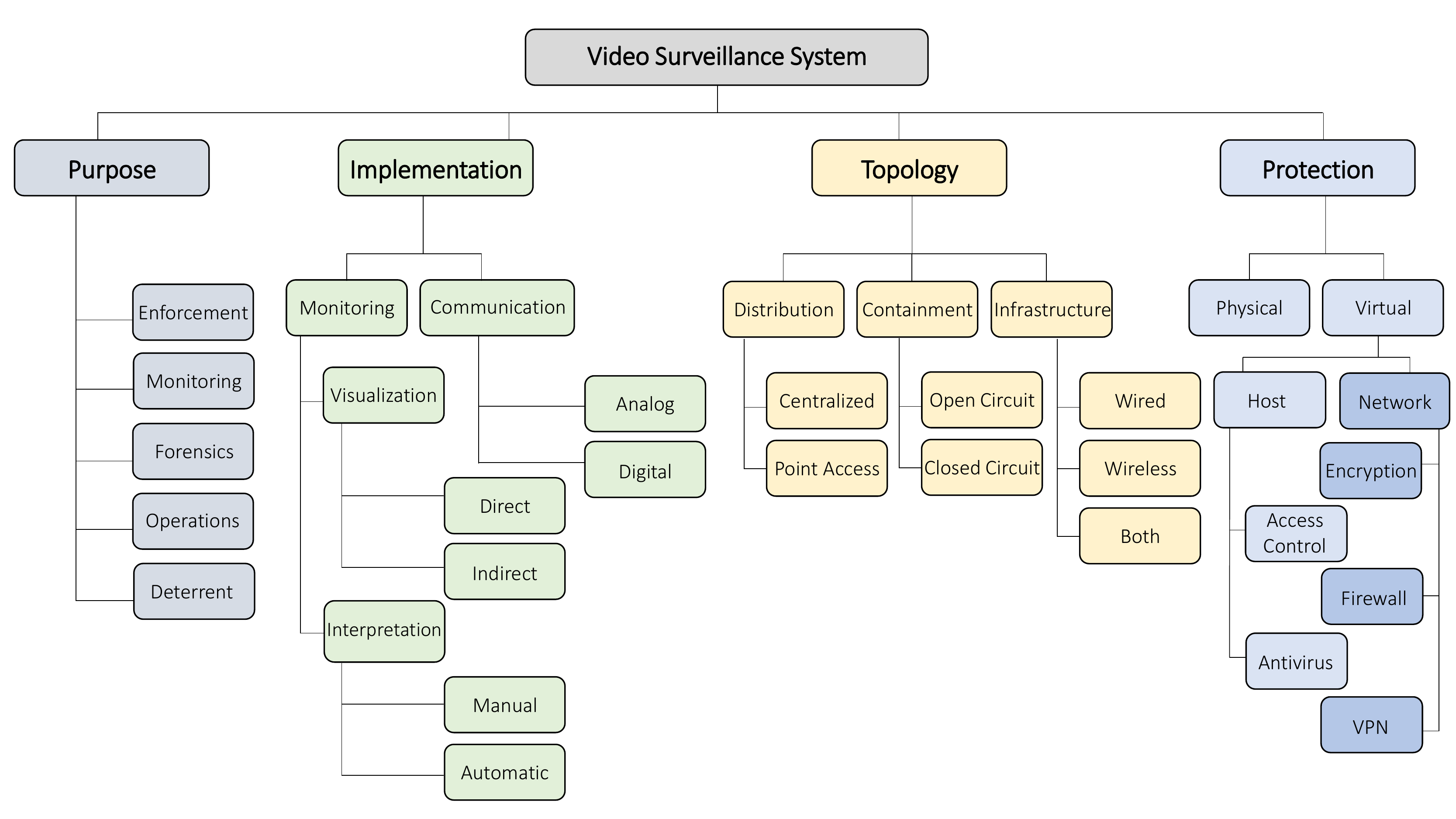}
	\caption{An Overview of Video Surveillance Systems}
	\label{fig:taxonomy}
\end{figure*}

To get a better understanding of IP-based video surveillance systems, in Fig. \ref{fig:taxonomy} we present an overview of their concepts. In general, a system can be described in terms of its purpose, implementation, topology, and protection. We will now detail each of these aspects.

\begin{description}
	\item[Purpose] The purpose of a video surveillance system depends on the user's needs. 
	\begin{description}
		\item[Enforcement] The user may want to send security forces or police to an area undergoing some violation of law or protocol. This is common in governments, transportation services, stores, and even workplaces.
		\item[Monitoring] The user may want to know what is happening in a certain location for some general purpose, or to have a sense of security. For example, home, baby, and pet monitoring.
		\item[Forensics] The user may want to be able to produce evidence or track down an individual.
		\item[Operations] The user may want to improve operations by having an overview of what is going on. For example, employees can be guided or managed more efficiently.
		\item[Deterrent] The user may want to have the system visually present to simply ward off potential offenders or trespassers. In some cases, the user will not even have a means of viewing the video footage. 
	\end{description}

	\item[Implementation] There are various ways the hardware/software of the system can be setup to collect and interpret the video footage. We categorize the system's implementation into two categories:
	\begin{description}
		\item[Monitoring] This concept regards how the user visualizes the video streams, and how the content is interpreted. The visualization can be provided directly to the user directly such as in a closed circuit monitoring station, or indirectly via a digital video recorder (DVR) with remote access or in the cloud. The interpretation of the content can be done manually by a human user reviewing the content, or automatically via motion detection, or advanced applications such as object tracking, image recognition, face-detection, and event-detection. 
		\item[Communication] The refers to the means in which the system transports the video feeds. With analog methods, the video is sent to the DVR as an analog signal (which is subsequently connected to Internet). With digital methods, the video is processed, compressed, and then sent as a packet stream to the DVR via IPv4 and IPv6 network protocols. A common approach is to compress the stream with the H.264 codec and then send it over the network with a real-time protocol such as RTP over UDP. 
	\end{description}
	
	\item[Topology] An IP-based surveillance system's topology can be described by its distribution, containment, and infrastructure. Distribution refers to whether the cameras are located anywhere in the world or physically located in one area. Containment refers to whether the system is closed circuit (not connected to the Internet) or open circuit --and relies on access control to deny users without proper credentials. Finally, infrastructure refers to how the elements of the system are connected together: wireless (e.g., Wi-Fi), wired (e.g., Ethernet via CAT6 cables), or both.
	
	\item[Protection] The protection of surveillance system refers to how the user secures physical and virtual access to the system's assets and services. Without physical protection, an attacker can tamper/damage the cameras or install his/her own equipment on the network. Virtual protection can be employed on the network hosts or on the network itself:
	\begin{description}
		\item[Host] Cameras, DVRs, and other devices can be protected by using proper access control mechanisms. However, like any computer, these devices are subject to the exploitation of un/known vulnerabilities in the software, hardware, or simply due to user misconfiguration (e.g., default credentials) \cite{antonakakis2017understanding}. Protecting the hosts from attacks may involve anti-virus software or other techniques.
		\item[Network] Depending on the topology, access to the system's devices may be gained via the DVR, an Internet gateway, or directly via the Internet. A user may protect the devices and the system as a whole by securing the network via encryption, firewalls, and end-to-end virtual private network connections (VPN).
	\end{description}
\end{description}

\subsection{Assets}\label{subsec:assets}
An asset is a thing of value which may be targeted by an attacker. In our case, the assets are data, devices, software, and infrastructure:

\begin{description}\label{subsec:UserCriticalAss}
	
		\item[DVR - Media Server] The digital video recorder, or other media server, which is responsible for receiving, storing, managing, and viewing the recorded/archived video feeds. DVRs are typically an application running on the user's server, or a custom hardware Linux box. DVRs can also be a cloud based server. In a small system, there may be cameras which do not support a DVR, and require the user to connect to the camera directly (e.g., via web interface).
		
		\item[Cameras] The devices which capture the video footage. There are many types, brands, and models of IP-Cameras, each of one has its own capabilities, functionalities, and vulnerabilities. For configuration, some IP cameras provide web-based interfaces (HTTP, Telnet, etc) while others connect to a server in the cloud. Most camera act as a web servers which provide video content to authorized clients (e.g., the DVR will connect to the camera as a client).
		
		\item[Viewing Terminal] The device/application used to connect to the DVR or camera in order to view and manage the video content. For example, an Android application running on a smartphone or the DVR itself.
							
		\item[Network Infrastructure] The elements which connect the cameras to the DVR, and DVR to the user's viewing terminal. For example, routers, switches, cables, etc. The infrastructure also includes Virtual Private Network (VPN) equipment and links. VPNs are LANs which tunnel Layer 2 (Ethernet) traffic across the Internet, between gateways and user devices, using encryption. Site-to-site VPNs can bridge two segments of a the surveillance network over the Internet. A remote-site connection tunnels traffic directly from a user's terminal to the surveillance network.
		
		\item[Video Content] The video feeds which are being recorded or that have been archived for later viewing.

		\item[User Credentials] The user names, passwords, cookies, and authentication tokens used to gain access to the DVR, cameras, and routers. The credentials are used to authenticate users and determine access permissions of video content, device configurations, and other assets.
		
		\item[Network Traffic - Data in Motion] Data being transmitted over the network infrastructure. This can be credentials, video content, system control data \cite{Liu2005} (e.g., pan, tilt, or zoom), and other network protocols (ARP, DNS, HTTP, SSL, TCP, UDP, etc).

\end{description}

\subsection{Deployments}
\begin{figure}
	\centering
	\includegraphics[width=\linewidth]{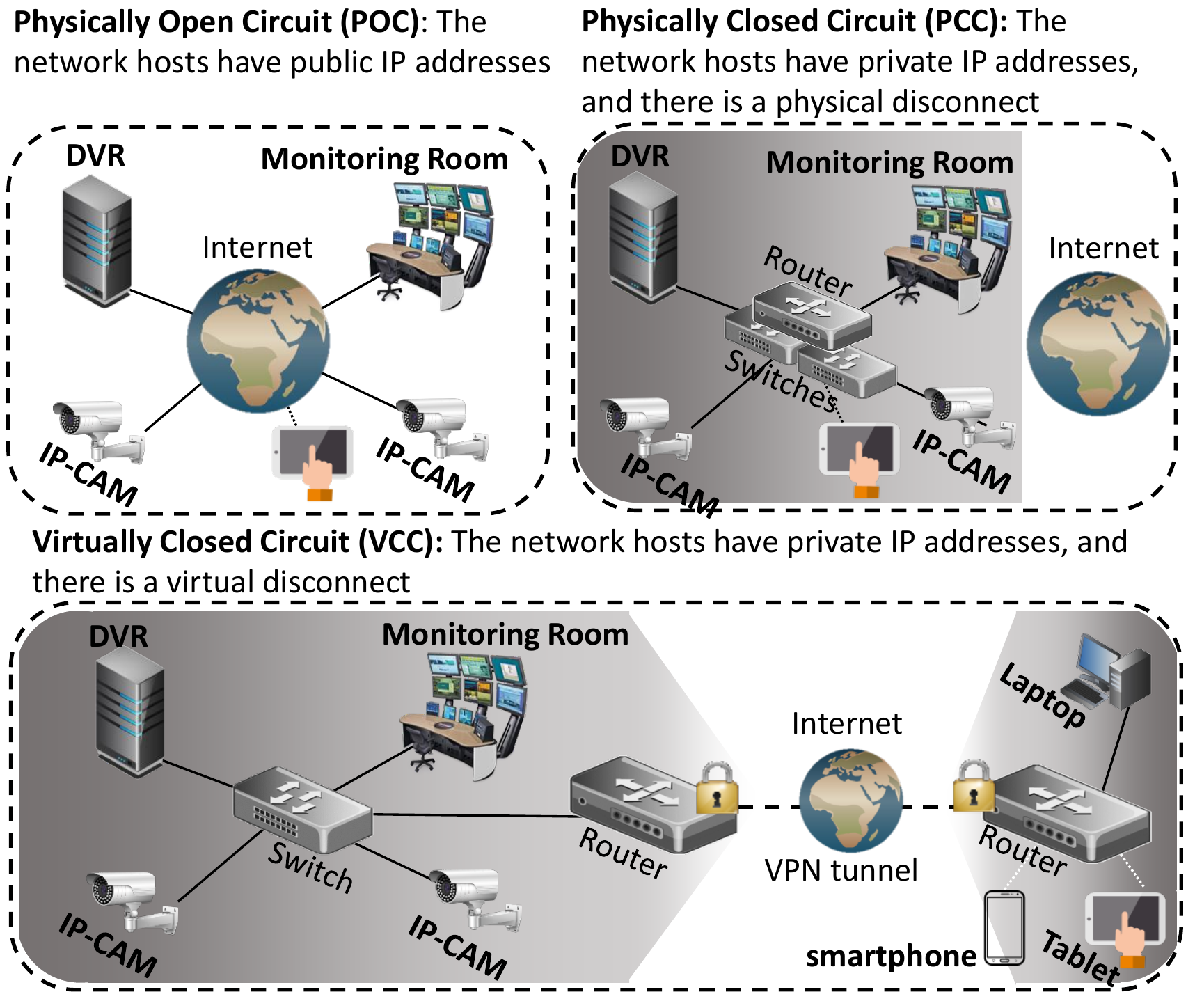}
	\caption{Accessibility models for deploying a video surveillance system.}
	\label{fig:Deployments}
\end{figure}

There are several ways an IP-based surveillance system can be deployed. The the network topologies can be centralized (all cameras connect to a DVR) or distributed (the user connects to each individual camera). In terms of accessibility, the system can being/directly accessible via the Internet, or not at all. In this regard, we identify three categories of accessibility (visualized in Fig. \ref{fig:Deployments}):
 
\begin{description}
	\item[Physically Open Circuit (POC)] When the network hosts in the system (cameras, DVR, etc) have public IP addresses. This means that anybody from Internet can send packets to the devices.
	\item[Physically Closed Circuit (PCC)] When the network hosts in the system have private IP addresses, and there is no infrastructure which connects the network to the Internet. This means that nobody from Internet can send packets to the devices directly. These systems are also called air-gapped networks \cite{guri2018bridgeware}.
	\item[Virtually Closed Circuit (VCC)] When the network hosts in the system have private IP addresses, and the network is connected via the Internet using a VPN. This means that nobody from Internet can send packets to the devices directly, unless they send packets via the VPN.
\end{description}

\section{Security Violations}\label{subsec:AttackerGoals}
A security violation can be described as an attack on the system's \textbf{c}onfidentiality, \textbf{i}ntegrity and \textbf{a}vailability (known as the \textit{CIA} triad). A violation of confidentiality refers to the unauthorized access of information.  A violation of integrity refers to the intentional manipulation and alteration of information. Finally, A violation of availability refers to the act of preventing authorized users from accessing services or resources when needed. 

The goals which an attacker may have in assaulting the system, can be described in terms of the CIA triad:

%

\begin{enumerate}
	\item \textbf{Confidentiality Violation} - the unauthorized access of video content, user credentials, network traffic. In this case, the attacker intends to observe the video footage for his/her own nefarious purposes. As a result, this goal puts the privacy and physical security of the premises at risk. 
	\item \textbf{Integrity Violation} - the manipulation of video content, or the active interference of a secure channels in the system (e.g., the POODLE SSL downgrade attack). In this case, the attacker intends to alter the video content (at rest or in transit). Alteration can include freezing frame, looping an archived clip, or inserting some other content. This misinformation can lead to physical harm or theft.

	An attacker may violate a system's integrity for a goal which is not directly related to the video content. For example, the attacker may want to exploit the system's vulnerabilities to gain \textit{lateral movement} to external assets. The system may be used as a stepping stone to gain access to the following external assets:
	
	\begin{enumerate}
		\item \textbf{Internal network} - surveillance systems (especially closed circuit systems) may be connected to the organization's internal network for management purposes. An attacker may leverage this link in order to gain access to the organization's internal assets.
		
		\item\textbf{Users} - users of the system may be targeted by the attacker. For example, the attacker may wish to install ransomware on the viewing terminal, or to hijack a the user's personal accounts. 
		
		\item \textbf{Recruiting a Botnet} - A `bot' is an automated process running on a compromised computer which receives commands from a hacker via a command and control (C\&C) server. A collection of bots is reffed to as a botnet, and are commonly used for launching DDoS attacks, mining crypto currencies, manipulating online services, and performing other malicious activities. An example botnet which infected affected IP-cameras and DVRs was the Mirai malware botnet. In 2016, the Mirai botnet generated a 1.1Tbps DDoS attack against websites, webhosts, and service providers. 
		%
	\end{enumerate}
	
	\item \textbf{Availability Violation} - the denial of access to stored or live video feeds. In this case, the attacker's goal is to (1) disable one or more camera feeds (hide activity), (2) delete stored video content (remove evidence), or (3) launch a ransomeware attack (earn money). For example, the attack on Washington DC's surveillance system in 2017 \cite{ransomecam}. 
\end{enumerate}

\section{Attacks}\label{sec:AttackModels}
\begin{figure}[b]
	\vspace{-.5cm}
	\centering
	\includegraphics[width=\columnwidth]{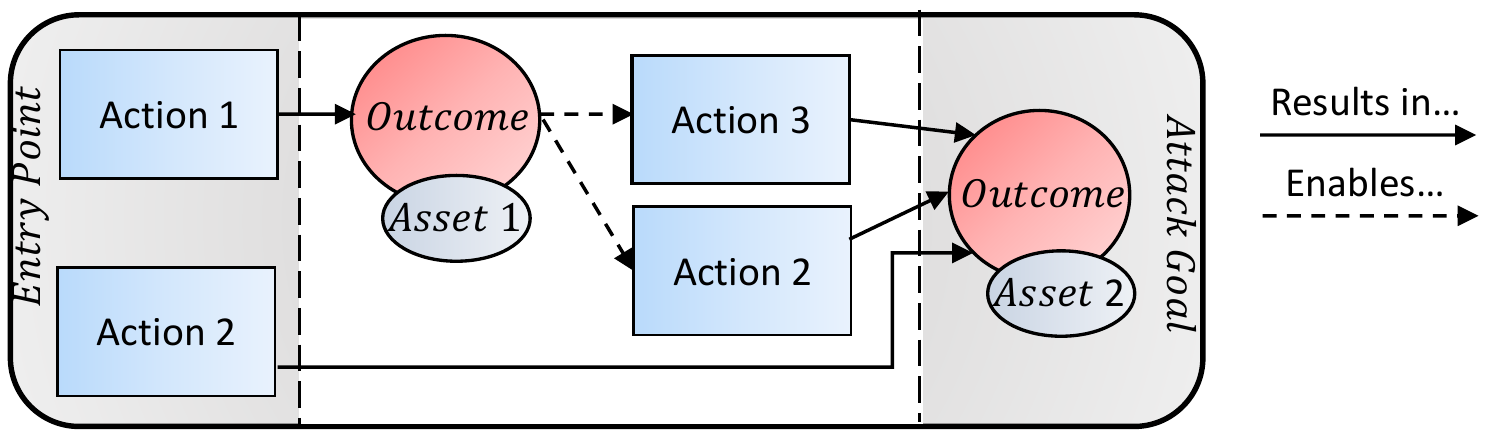}
	\caption{An example of two attack vectors which arrive at the same goal.}
	\label{fig:AttackModel}
\end{figure}

There are many different kinds of attacks. Some scenarios involve a single step (e.g., DDoS a VPN link), while others have numerous steps (e.g., stealing credentials by sending a phishing email, then installing a malware, and so on). A sequence of attack steps is often referred to as an attack vector. Each step in the vector gives the attacker access to some asset, and the final step in the vector fulfills the attacker's goal. As an example, Fig. \ref{fig:AttackModel} illustrates two attack vectors which arrive at the same goal: Attack 1 achieves an outcome which compromises an asset. As a result, the attacker can either perform Attack 2 or 3 to achieve the goal. Alternatively, the attacker may perform Attack 2 with directly achieves the goal in one step, but may be more difficult to accomplish.

To understand an attack vector, one needs to investigate the following aspects:
\begin{description}
\item[Threat Agent/Actor.] The person, device, or code which performs an attack step on behalf of the attacker.
\item[Threat Action.] The malicious activity which an agent can perform at each step (access, misuse, modify, etc.)
\item[Threat Consequence/Outcome.] What the attacker obtains at the successful completion of an attack step.
\item[Attack Goal.] The ultimate outcome which the attacker is trying to achieve (at the end of the attack vector).
\end{description}

We will now discuss each of these aspects with regards to the surveillance system.

\subsection{Threat Agents}\label{subsec:theatAgents}
We identify the following relevant threat agents/actors:

\begin{enumerate}
	\item \textbf{Hacker} - An individual who is experienced at exploiting computer vulnerabilities, whose  unauthorized activities violate the system's security policies. A Hacker can be in a remote location (i.e., the Internet) or in close proximity of physical network.
	
	\item \textbf{Network Host} - A computer connected to the system's network which is executing malicious code. The computer can be an IP-camera, DVR, or any programmable device in the network. A network host can become a threat actor via local exploitation (the un/intentional instillation of malware -- social engineering \& insiders), remote exploitation (e.g., exploit a web server vulnerability or an open telnet server), or a supply chain attack.

	\item \textbf{Insider} - An authorized user of the system who is the attacker or colluding with the attacker. The insider may be a regular user (e.g., security officer), an IT support member, or even the system's administrator. An insider may directly perform the entire attack, or enable a portion of an attack vector by installing malware, changing access permissions, etc.  
\end{enumerate}

\subsection{Threat Actions}\label{sec:attacks}
We identify the following assaults which a threat agent can perform on the system.

\subsubsection{Performing Code Injection} Code injection is an exploitation of improper parsing of an input which results in the input being executed as code. A threat agent may perform a code injection to reveal sensitive information or install some malware. One example is the cross-site request forgery (CSRF) flaw which can be used to add a secondary administrator account on some cameras (for example CVE-2018-7524, CVE-2018-7512). Another example is the SQL injection attack which have affected Geutebruck G-Cam/EFD-2250 cameras (CVE-2018-7528). Finally, stack-based buffer overflow vulnerabilities have been exploited, and have been discovered in IP-cameras and DVRs (CVE-2017-16725).

Some cameras, run local HTTP web servers to provide users with a convenient configuration interface. However, these servers may be outdated and vulnerable to attack, such as the infamous Heartbleed vulnerability in OpenSSL. Another example is the Sony surveillance camera IPELA series, where parsing vulnerabilities can be exploited to perform a buffer overflow attack via a simple HTTP post message (CVE-2018-3937/8). Other attacks on web servers found on IP-cameras include directory traversals and cgi-bin script exposures. Crafted URLs sent to the server can cause directory traversals which may reveal administrator and Wi-Fi credentials (CVE-2013-2560). Sending various inputs to exposed cgi-bin script URLs can enable live video feeds and enable telnet communications \cite{cgitelnet}.

\subsubsection{Manipulating/Observing Traffic} A threat agent may manipulate, reroute, or observe network traffic. For example, an agent may (1) perform a man-in-the-middle (MitM) attack in the local network, and then (2) freeze a video image or injected into a live feed. For the MitM attack, the attacker could reroute traffic through him via ARP poisoning, DHCP/DNS spoofing. For injection, the tool VideoJak may be used to exploit unencrypted video streams using the RTSP or RTP protocols. These protocols are commonly used in video surveillance systems, and may be left unencrypted if found in a PCC deployment. 

In the case of traffic observation, an agent may be able to observe video content. In \cite{tekeoglu2015investigating} the authors succeeded in extracting JPEG images generated by NetCam IP Camera by sniffing the network traffic. Furthermore, even when the video stream is encrypted, the video footage can be inferred by observing the stream's bandwidth patterns \cite{schuster2017beauty}. This is due to how video codecs (such as H.264) compress motion between frames, and how clients buffer content.
Moreover, observing traffic can also reveal network topological information from universal plug and play (UPnP) traffic, and credentials may be revealed as plain text in HTTP traffic (e.g., the DVR in CVE-2017-15290). 

\subsubsection{Exfiltrating Information}
Cameras can be exploited to exfiltrate information for an attacker \cite{costin2016security}. For example, a malware contained within an isolated network can blink an LED light in view of the camera which is connected to the Internet. By modulating the blinking pattern, the attacker can ex filtrate some stolen information (e.g., user credentials) to a remote location.
		
\subsubsection{Flooding \& Disrupting} A threat agent may prevent access to a service or data by send flooding the network with packets, or sending crafted traffic to a network application. A classic DoS attack is to flood a DVR or Camera until the server's resources are depleted and all new (and sometimes existing) sessions are blocked (e.g., CVE-2019-6973). For example, using the hping3 tool, a TCP SYN flood can disable a web server (e.g., CVE-2018-9158), a UDP flood can overload a network interface, and an ISKAMP flood can disable a VPN connection. Furthermore, an SSDP amplification attack can be used to overload a DVR. In this attack, the agent causes the cameras to spam the DVR with large amounts of UPnP meta-data by sending requests using the DVR's IP address. 

IP-cameras are often susceptible to these attacks because they are typically resource limited devices. For example, some cameras can only support up to 80 concurrent HTTP connections, which can easily be consumed. Another example an SSL regeneration attack where the agent repeatedly requests key renegotiations which overloads the device's CPU. 

Other DoS attacks can be accomplished by exploiting bugs and vulnerabilities. For example, a camera can be crashed by sending large HTTP POST requests (CVE-2018-6479), and a VPN router can be forced to drop all connections due to crafted packets (CVE-2014-0674 and CVE-2016-6466). 

\subsubsection{Scanning \& Reconnaissance} A threat agent may perform a network scan to learn the topology, assets, open network ports, and services available for potential exploitation. Off the shelf tools such as NMAP can be used to map the network and reveal information about its hosts. An agent may also elicit responses from web services to reveal version information, and perform fuzzing attacks on exposed web interfaces to find potential vulnerabilities. Fuzzing is typically performed off-site since it is easy to detect.

\subsubsection{Exploiting a Misconfiguration} A threat agent may utilize a misconfiguration to install malware or gain access to sensitive data. Example misconfiguratiosn include default credentials, exposed services (e.g., Telnet), and improper access control rules. A misconfiguration can be caused by a user of the system or the manufacturer.

\subsubsection{Performing a Brute-Force Attack} A brute-Force attack is the attempt of guessing a correct input by trying many possible options. Brute-Force attacks can be used to reveal user credentials such as user names and passwords. These attacks can be mitigated by limiting the number of failed logins allowed per minute. However, in some cases, camera manufactures do not implement this security feature. To arrive at a solution quickly, a dictionary of common passwords may be used as a guessing pool. For example, the \textit{Mirai} malware propagated to other devices by connecting via Telnet using a dictionary of 62 common credentials used by cameras, DVRs, and IoTs alike \cite{antonakakis2017understanding}. Another example, is the \textit{Remaiten} and \textit{Aidra} malwares which compromised cameras and other IoT devices using a similar approach \cite{Liu2005}.

\subsubsection{Social Engineering} Social Engineering (SE) refers to psychological manipulation of a person which causes him/her to perform an action on behalf of the attacker.  
	Common SE attacks include phishing emails and baiting. In phishing, the threat agent sends a message (email, SMS, etc) disguised as trustworthy source, in an attempt to get the receiver to install some malware, or ultimately reveal user credentials. In baiting, the threat agent plants a multimedia device (e.g., USB drive or microSD card) loaded with malware. The victim then unwittingly plugs it into his machine which infects it. 
	

\subsubsection{Physical Access} Physical access is where a threat agent performs an attack which requires direct physical contact with the system. For example, installing a wiretap, backdoor device, accessing a terminal in the server room, flashing a camera's firmware, obstructing the camera's view, or simply cutting a wire. 

\subsubsection{Reverse Engineering} A threat agent may learn the target device's credentials or vulnerabilities by the use of reverse engineering (RE). Reverse engineering is typically performed off-site using the same hardware/software used by the victim. RE is the process of analyzing compiled code or hardware to identify system's components and their interrelationships. During this process, vulnerabilities and even hard-coded credentials can be discovered. 

One approach is to analyze the pre-compiled firmware provided by the manufacturer. In a Black Hat lecture \cite{Exploiting_Surv_cams} the authors focused on IP cameras which face the Internet and analyze them through firmware images supplied by the camera's vendors. The authors found zero-day vulnerabilities in digital surveillance equipment from various firms including D-Link Corp, Cisco Systems, Linksys, TRENDnet, and more with the use existing tools. The analysis revealed serious security vulnerabilities such as administrative passwords, remote code execution vulnerabilities, and more. Another case was found in Sony's IPELA surveillance camera series. By performing RE on the firmware, researchers from Sec Consult found a backdoor via two hard coded root level credentials. These backdoors have also been discovered in other cameras and DVRs (In CVE-2018-5723 and CVE-2017-6432). The hard-coding of credentials may occur intentionally, or by mistake (e.g., a developer forgot to remove the credentials after testing).

Another approach of RE is to interface with the device via its Universal Asynchronous Receiver/Transmitter (UART) ports. These ports are typically inside the device's casing, and used by the manufacturer for debugging purposes. UART ports can be used to expose vulneabilites, gain access to the firmware, run foreign applications, extract sensitive information, or upload custom firmware for further analysis.	

\subsubsection{Adversarial Machine Learning} \label{subsubsec:adv}
Video surveillance requires either a manual or automated way of reviewing the video content for events. For example, detecting live intrusions or locating suspects. Therefore, the domain of video analytics have been applied to minimize the human efforts in this task. In the case of large deployments, such as China's state surveillance system, (see https://www.nytimes.com/2018/07/16/technology/china-surveillance-state.html) automated methods are required. Some automated technologies include, facial recognition \cite{ding2018editorial}, event detection \cite{li2015crowded}, and object tracking \cite{joshi2012survey}. However, since most of these technologies rely on machine learning, they are susceptible to adversarial attacks \cite{akhtar2018threat}. An adversarial attack is where a machine learning model is abused by either (1) poisoning the model during training so that the mode will behave according to the attacker's will, (2) crafting an input which will yield an unexpected output, or (3) learning the training data or the model itself by observing the input-output relationship. Adversarial attacks on these technologies mean that an attacker may be able to evade detection, falsify the recognition of an object, or even cause a DoS attack by raising the technology's false positive rate.

\subsection{Threat Consequence}\label{subsec:attackOutcomes}
The success of an assault during an attack step provides the attacker with new capabilities. For example access to new assets, the ability to run code, and the ability to perform new attacks. We identify the following as the primary threat consequences:

	\subsubsection{Privilege Escalation} An attacker may receive new credentials or execute code in a way that provides access to previously restricted assets. This escalation can be used to gather information, un/install software, en/disable a protection mechanism, etc. For example, an unprotected web facing CGI method can give an unauthenticated user the ability to bypass the login screen and access the webcam contents including: live video stream, configuration files with all the passwords, system information, and much more (CVE-2017-17101). Another example is CVE-2017-6432 where one can inject new users into DVR management traffic via a MitM attack.

    \subsubsection{Access to Video Footage} The attacker may be able to watch/download live or pre-recorded video footage.

	\subsubsection{Arbitrary Code Execution (ACE)} A significant security threat which enables an attacker to execute any command on a target machine or within a target process. As a result, the attacker can perform privilege escalation, install malware, steal data, and perform other malicious tasks. ACE vulnerabilities have been discovered on IP cameras, DVRs and VPN routers(for example: CVE-2018-6414, CVE-2018-9156, CVE-2018-9157, CVE-2018-7532, CVE-2018-7512, CVE-2015-8039, CVE-2018-0125, CVE-2017-3882). 
	
	\subsubsection{Installation of Malware}	The attacker may be able to install and execute his own process on a target device. This software is referred to as malware: malicious code designed to damage a computer with malicious intent. Types of malware include worms, trojan horses, viruses, spyware, scarewares, launchers, ransomeware, adware, and rootkits. Malware can be used to steal sensitive data, encrypt or delete user data, harm the device, mine crypto currencies, add the device to botnet, or act as a pivot point for lateral movement through the victim's network. 
	
	\subsubsection{Lateral Movement} An attacker may again a stronger foothold in the surveillance system, and achieve the ability to reach previously inaccessible assets. The attacker may also be able to reach other systems and infect user devices connected to the system.
	
	\subsubsection{Man in the Middle (MitM)} An attacker may be able to covertly observe and manipulate traffic between two or more endpoints. A MitM can harm the confidentiality, integrity, and availability (CIA triad) of the system. For example, the MitM can eavesdrop, manipulate, craft, or drop network traffic.
	
	\subsubsection{Denial-of-Service (DoS)} An attacker may be able to affect the availability of a service, data, or resource. If the attacker has compromised cameras or the DVR, then attacker can cause a camera stop transmitting video content, delete historic content, block access to the DVR, or cause a VPN link to fail. As a result, a crime may be accomplished on premises without digital evidence. An attacker may also be able to evade detection without raising any alerts in the DVR. This can be accomplished via a video injection attack or an adversarial machine learning attack. 
	
	With regards to DDoS attacks: an attacker may target the surveillance network with a remote botnet. In this case, the consequence of not filtering the traffic is a DoS to the system. We also note that in the case where system itself is infected with a botnet, and is then used to launch a remote DDoS attack, the consequence may still be a DoS to the local system since there will be congestion and the ISP may block the system from network access.

	\subsubsection{Access to an Isolated Network} In some cases, the DVR is connected to the Internet (POC or VCC) and is also connected to a network which is supposedly isolated the Internet (e.g., airports, hospitals, factories, etc.) By compromising the DVR or one of the cameras, the attacker can perform lateral movement into the isolated network. For example, two researchers hacked into a Google office building's air-conditioning system portal and then gained access to the internal network \cite{Research15:online}.

\subsection{Example Attack Vectors}\label{subsec:attackVector}
In this section we provide example attack vectors for different scenarios. Although there are many possible attack vectors, we will illustrate a small sample of common vectors used to attack IP-based surveillance systems. For the illustrations, we use the template presented in Fig. \ref{fig:AttackModel}.

\begin{figure}[ht]
	\centering
	\begin{subfigure}[t]{\columnwidth}
		\caption{Unauthorized Video Monitoring: POC Deployment}
		\centering\includegraphics[width=\textwidth]{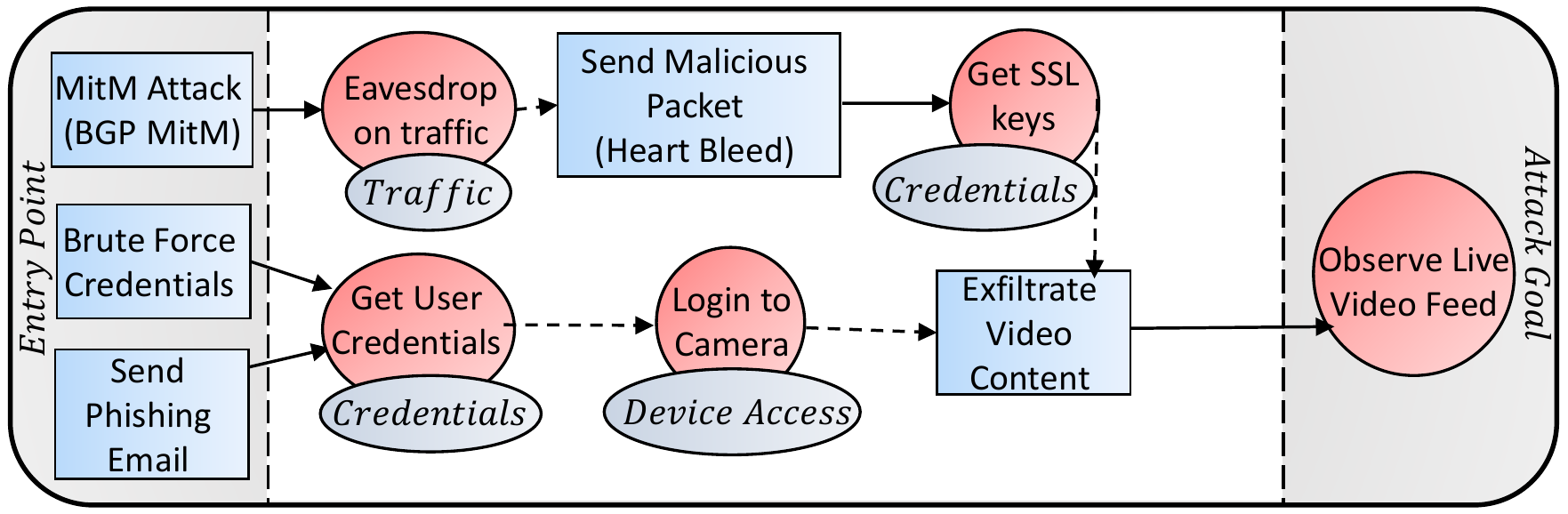}
		\label{fig:av_1}
	\end{subfigure}\\%
	\begin{subfigure}[t]{\columnwidth}
		\caption{Stealing Archived Video Footage: POC Deployment}
		\centering\includegraphics[width=\textwidth]{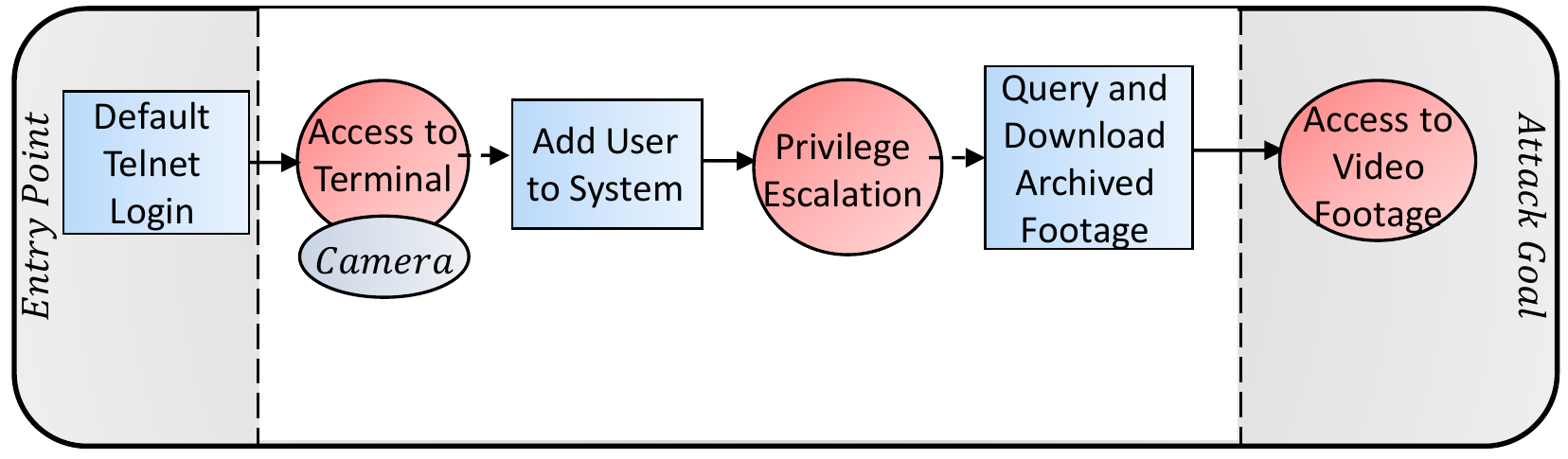}
		\label{fig:av_4}
	\end{subfigure}
	\begin{subfigure}[t]{\columnwidth}
		\caption{Accessing an Air-Gapped System: PCC Deployment}
		\centering\includegraphics[width=\linewidth]{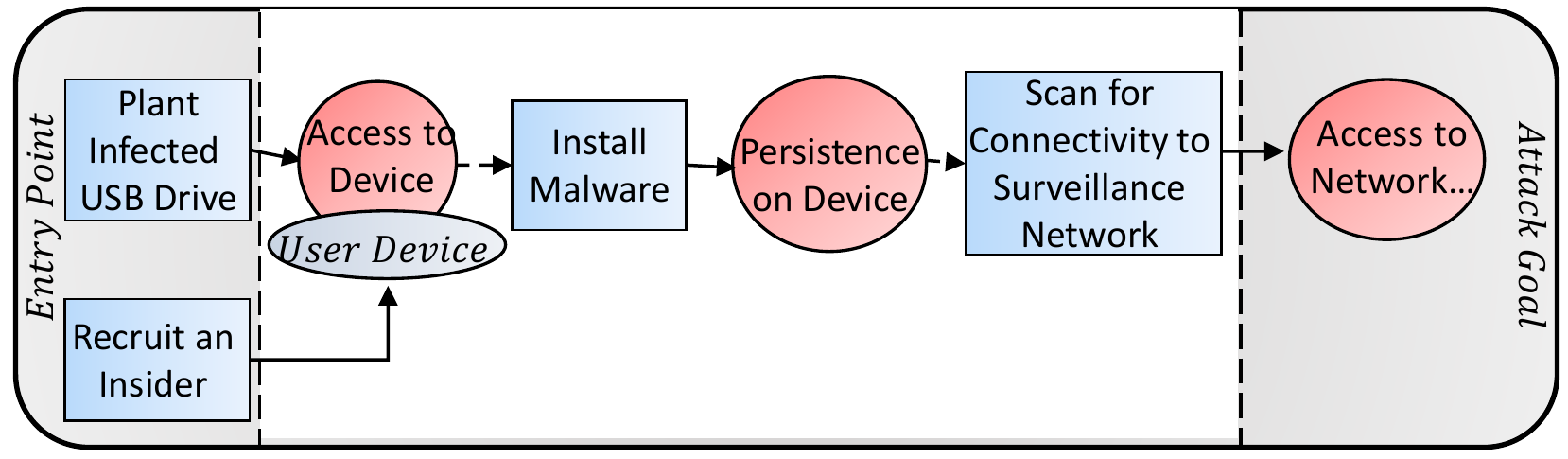}
		\label{fig:av_3}
	\end{subfigure}
	\begin{subfigure}[t]{\columnwidth}
		\caption{Disabling Video Feeds: VCC Deployment}
		\centering\includegraphics[width=\textwidth]{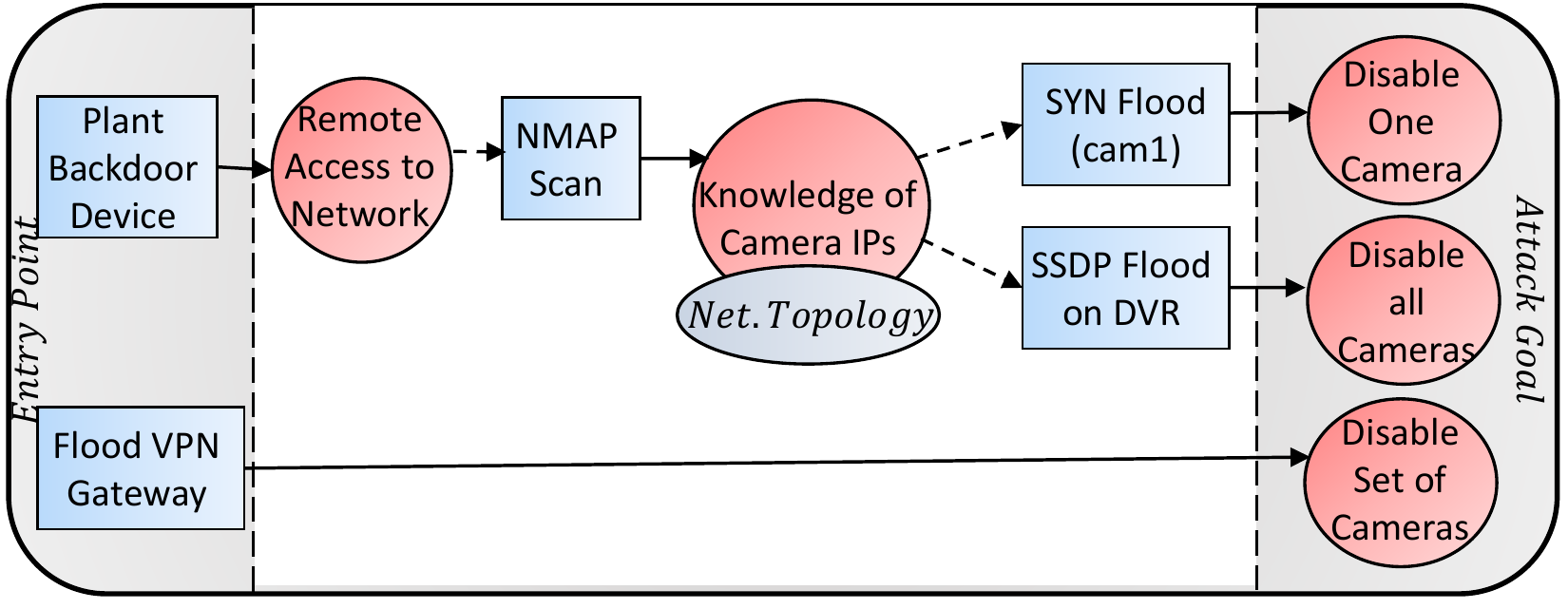}
		\label{fig:av_2}
	\end{subfigure}

	\caption{Example attack vectors on IP-based surveillance camera systems deployed with POC, PCC, and VCC topologies.}\label{fig:vectors}
\end{figure}

\subsubsection{Unauthorized Video Monitoring}
Consider an attacker who wants to view the video footage of a POC deployment with encrypted traffic. In Fig. \ref{fig:vectors}\subref{fig:av_1}, a few potential attack vectors are illustrated. A state actor may perform a BGP MitM routing attack, and cause all of the video surveillance traffic to pass through them first. Next, the attacker may exploit the Heartbleed vulnerability to get the SSL cryptographic keys and then decrypt the video traffic. A simpler way might be to get the camera's or DVR's login credentials by performing a brute-force login attack, or to send phishing emails to users of the system to have them unwittingly reveal their credentials.

\subsubsection{Stealing Archived Video Footage}
Another scenario is the case where an attacker wants to blackmail an individual by obtaining sensitive video footage. Fig. \ref{fig:vectors}\subref{fig:av_2} we illustrates one possible attack vector where an agent gains access to the DVR's terminal by performing a dictionary brute-force login attack on the DVR's telnet server. Next, the attacker retrieves the desired footage and exfiltrates it out of the network.

\subsubsection{Accessing an Air-Gapped System}
In the case of a PCC deployment, direct access from the Internet is impossible. However, this does not meant that the network is impervious to infiltration. In Fig. \ref{fig:vectors}\subref{fig:av_3}, we illustrate how an attacker can install malware on one of the user devices, such as a viewing terminal (tablet, console, etc.) This can be accomplished by surreptitiously placing an infected USB drive in the area, or by recruiting an insider. Next, the malware will be installed by the threat agent, which will then subsequently can and connect to the surveillance network. At this point, the malware may perform automated actions designed by the attacker (e.g., disable camera at a certain time) or it may be able to communicate with the attacker directly via bridgeware \cite{guri2018bridgeware}.

\subsubsection{Disabling Video Feeds}
An attacker can disable video feeds in many different ways. Let's assume that the target system has a VCC deployment, so access is either physical or via a VPN gateway. In Fig. \ref{fig:vectors}\subref{fig:av_4}, we show how an attacker can disable one, all, or a subset of cameras in this scenario. First, an attacker may plant a backdoor device to gain remote entry (e.g., under the pretext of a repairmen, the attacker secretively connects a raspberry Pi to the network, and then connecting to the Pi's Wi-Fi access point). Next, the attacker will have the device, scan the network to reveal the IP addresses of the cameras and the DVR. Finally, single cameras are disabled via TCP SYN flooding, or all cameras are disabled by exploiting a potential SSP flood vulnerability in the DVR. Alternatively, instead of planting a backdoor device, the attacker may be able to perform a flood attack (e.g., ISAKMP flood) 
on a site-to-site VPN gateway which will result in a set of nearby cameras to go off-line.

\section{Countermeasures \& Best Practices}\label{sec:countermeas}
In the following section we review existing countermeasures and best practices which can be used to protect modern surveillance systems. 

\subsubsection{Intrusion Detection \& Prevention Systems} 
Basic cyber defense should be considered in every computer network. For example, to detect and prevent malware infections, anti-virus software should be installed on the user terminals and DVRs. In non-distributed POC topologies, a strict firewall should be deployed to pass the minimal network traffic required to use the system (e.g., block telnet, ICMP `ping' packets, etc). Encase the adversary evades the firewall, a network intrusion detection system (NIDS) can be used to detect malicious traffic patterns. In this case, free NIDS such as Snort and Suricata, or commercial software, can be used.

\subsubsection{Configurations \& Encryption} 
One should carefully review the configurations of the cameras, routers, terminals, and DVR. For example, weak or default passwords should be changed, and different passwords should be used among different devices if possible. Moreover, APIs and other similar features should be disables if not needed. It is also important to enable secure communication wherever possible, and to ensure that devices are not using self-signed SSL certificates (a common default setting). Finally, one should periodically check for new CVEs that the software/firmware of all devices are up to date.

\subsubsection{Restrict Physical Access} 
The most basic perimeter defense is to restrict physical access to the system's assets. If possible, wiring should not pass through public areas, all networking equipment (switches, routers, etc.) should be protected under lock-and-key, and access to the system should be managed, logged, and monitored.  

\subsubsection{Defense against DoS attacks} 
There are many protocols and vulnerabilities  can be abused to perform a DoS attack. As a result, there are many different defense mechanisms which can be deployed. Good protection involves the following steps: (1) detect the attack's initiation, (2) select the malicious/harmful packets, and (3) filter/log the detected packets. For the attack detection, machine learning and statistical methods can be used --such as light weight anomaly detection. 

\subsubsection{Defense against MitM Attacks} 
Proper encryption should be used to prevent eavesdropping and packet manipulation (e.g., injecting video) as a result of a MitM attack. However, sometimes vulnerabilities are discovered in encryption protocols, and systems may be misconfigured. Therefore, as an additional line of defense, additional methods can be deployed. To detect tampering (video injection), one can reference time according to shadow positions. However, this method only works in limited circumstances. Another method is to perform \textit{watermarking} \cite{bala2015brief}: a process of embedding a digital code into digital video sequences. When the video content is altered the watermark is significantly affected which alerts the receiver. To detect an eavesdropper or the presence of a third party with the ability to manipulate traffic, one can perform echo analysis to detect when malicious parties intercept traffic \cite{mirsky2018vesper}.

\subsubsection{Education} 
In many advanced persistent threats (APT) the initial intrusion comes in the form of a social engineering attack. The most effective way of mitigating these initial incursions, is to: (1) educate the users of the system of the potential attack vectors, and (2) warn users to be careful of unsolicited messages and requests made under false pretexts.

\section{Discussion}
We have identified two main emerging threats to IP-based video surveillance systems.
The first is \textit{adversarial machine learning} (Section \ref{subsubsec:adv}).
Advanced machine learning techniques, mainly based on deep learning, are being researched and integrated within today's video surveillance systems for automating various tasks including: weapon detection~\cite{castillo2019brightness}, fire detection~\cite{muhammad2018convolutional}, face recognition~\cite{ding2018trunk}, and anomaly detection~\cite{ribeiro2018study}.
In parallel, there has been an increase of research on adversarial machine learning~\cite{akhtar2018threat} meaning that these systems are vulnerable to attacks~\cite{evtimov2018robust}.
The second emerging threat is \textit{how these systems are being infected and recruited into botnets}, leading to attacks on the internal network (e.g., data exfiltration, spying or using the surveillance system for lateral movement) or on other external networks (e.g., DDoS, SPAM).

To address these threats, future work should focus on protecting the machine learning algorithms and by making them more resilient to adversarial machine learning attacks~\cite{goswami2018unravelling}, and by developing targeted security solutions for video surveillance systems to improve their protection against cyber-attacks.

New attacks are constantly emerging. 
As a result, a recent research trend for securing surveillance systems has been the use of advanced anomaly detection. 
With anomaly detection researchers are able to identify man-in-the-middle-attacks~\cite{mirsky2018vesper}, video injection, OS fingerprinting, fuzzing and ARP poisoning attacks~\cite{mirsky5kitsune}, and DDoS attacks~\cite{meidan2018n}.

Updating the software of such systems is also a challenging task since manufactures are focused on their next product, and in many cases do not have the capability of performing remote patching.
Therefore, we believe future research should focus on providing an external continuous protection that can be easily updated with information on newly discovered attacks.
One way to collect intelligence on emerging threats to surveillance systems is to use an advanced honeypot system~\cite{tambe2019detection}.
Moreover, by identifying emerging exploits, administrators can protect their systems before they get infected.

Finally, although in most cases the communication of advanced video surveillance systems is encrypted, the confidentially of entities can be compromised using side channel attacks as was shown by Nassi et al.~\cite{nassi2018game}.
Therefore, future research should focus on detecting and eliminating side channels.



\color{black}
\section{Conclusion}\label{sec:conclusion}
In this survey, we have reviewed the security of modern video surveillance systems. We have presented an overview of these systems, presented common deployments, and listed the system's assets. Using this information, we then reviewed the security of these systems by exploring the system's attack surface, enumerating the attacker's capabilities, and by providing some example attack vectors. Finally, we provided a concise summary of best practices and security solutions which can be used to enhance the security. We hope that this article will aid in securing existing and future video surveillance systems.

%
%
%

\bibliographystyle{unsrt}
\bibliography{IP_camera_survey}

\begin{IEEEbiographynophoto}{Naor Kalbo} is M.Sc. student at the Software Information-System Engineering department and a researcher at the Cyber Security Research Center (CSRC) at Ben Gurion University (BGU). His research lies in the field of attacks detection in the cyber-space domain using machine-learning, more specifically, detection of Man-in-the-Middle attacks through timing analysis. Naor has also used as a TA in ‘computer and network security’ course in the academia. Outside academia, Naor is cyber-security and ML enthusiasts with several years of experience in the field of system’s
development and researching at the industry.
\end{IEEEbiographynophoto}
\vspace{-.8cm}

\begin{IEEEbiographynophoto}{Yisroel Mirsky}
	received his M.Sc. (2016) in Information System Engineering from Ben-Gurion University (BGU). He is currently perusing a Ph.D. at BGU and is a research project manager in the BGU Cyber Security Research Center. His research interests and publications cover the topics of time-series anomaly detection, network intrusion detection, smart phone security, IoT security, and air-gapped networks.
\end{IEEEbiographynophoto}
\vspace{-.8cm}

\begin{IEEEbiographynophoto}{Asaf Shabtai}
	holds a B.Sc. in Mathematics and Computer Science (1998); B.Sc. in Information Systems Engineering (1998); a M.Sc. in Information Systems Engineering (2003) and a Ph.D. in Information Systems Engineering (2011) all from Ben-Gurion University. He is currently an assistant professor at the Department of Software and Information Systems Engineering at Ben-Gurion University. Asaf is a recognized expert in information systems security. His main areas of interest are cyber security and machine learning.
\end{IEEEbiographynophoto}
\vspace{-.8cm}

\begin{IEEEbiographynophoto}{Yuval Elovici}
	is the director of the Telekom Innovation Laboratories at Ben-Gurion University of the Negev (BGU), head of BGU Cyber Security Research Center, and a Professor in the Department of Software and Information Systems Engineering at BGU.	He holds B.Sc. and M.Sc. degrees in Computer and Electrical Engineering from BGU and a Ph.D. in Information Systems from Tel-Aviv University. His primary research interests are computer and
	network security, cyber security, web intelligence, information warfare, social network	analysis, and machine learning.
\end{IEEEbiographynophoto}
\vspace{-.8cm}

\end{document}